\documentclass{article}

\newcommand{\ket}[1]{|{#1}\rangle} 

\begin{document} 
\markboth{P.~A.~Zizzi}
{Qubits and Quantum Spaces}
%
%
\title{Qubits and Quantum Spaces}
\author{P.~A.~Zizzi\\
Dipartimento di Matematica Pura ed Applicata, Universita di Padova,\\ 
Via Belzoni 7, 35131 Padova, Italy\\
zizzi@math.unipd.it}
\date{}
\maketitle

\begin{abstract}
We consider the quantum computational process as viewed by an insider observer: this is equivalent to an isomorphism between the quantum computer and and a quantum space, namely the fuzzy sphere. The result is the formulation of a reversible quantum measurement scheme, with no hidden information.
\end{abstract}

\section{Introduction}

Performing the standard quantum measurement of a qubit in a classical world is equivalent to make a projection of the North or of the South pole of the Bloch sphere, into the 3-dimensional Euclidean space, each pole corresponding to one of the two computational basis states. The corresponding geometry (one point in 3-dimensional Euclidean space) has no rotational symmetry left, and this prevents to make any rotation of the Bloch sphere and recover the hidden quantum information. The irreversibility of a standard quantum measurement is due to the background geometry, which is commutative (classical). If instead, the background geometry is non-commutative \cite{con}, the North and South poles of the Bloch sphere are smeared out over the surface of a fuzzy sphere \cite{mad}; they are replaced by two cells. As such a quantum space is Lorentz-invariant, the corresponding measurement is a reversible operation as it is possible to recover the whole quantum information of the qubit by a rotation of the Bloch sphere. If, in particular, one considers a 2-points lattice, which is a subspace \cite{mar} of the fuzzy sphere, then, one is able to perform a projection of the North and South poles at the same time. The corresponding measurement, which we will call ''basic measurement'', is a linear superposition of two orthogonal projectors, and is a reversible operation since the 2-points lattice, although being a commutative geometry, allows rotations of the Bloch sphere about the z-axis.

\section{Irreversible and Reversible Measurements: The Background Geometries}

Let us consider a qubit in the superposed state:
\begin{equation}
    \ket{\psi}=a\ket{0}+b\ket{1}
\end{equation}
where $\ket{0}$ and $\ket{1}$ are the computational basis, and $a$ and $b$, called probability amplitudes, are complex numbers such that the probabilities sum up to one: 
$|a|^2+|b|^2=1$.
In vector notation we have: $\ket{0}\equiv\left(\begin{array}{c}1\\0\end{array}\right)$,
$\ket{1}\equiv\left(\begin{array}{c}0\\1\end{array}\right)$.
The standard quantum measurement of the qubit $\ket{\psi}$
in (1) gives either $\ket{0}$ with probability $|a|^2$, or $\ket{1}$ with probability $|b|^2$. 
This is achieved by the use of the two projector operators: 
$P_0=\left(\begin{array}{cc}1&0\\0&0\end{array}\right)$ and 
$P_1=\left(\begin{array}{cc}1&0\\0&1\end{array}\right)$.
The action of the two projectors on the superposed state (1) is, respectively:
$P_0\ket{\psi}=a\ket{0}$ and $P_1\ket{\psi}=b\ket{1}$.
After quantum measurement, the state (1) is left either in the state 
$\ket{\psi'}=\ket{0}$ or in the state $\ket{\psi'}=\ket{1}$.
Then, a lot of quantum information, that was encoded in (1), is made hidden. As a projector is not a unitary transformation, the standard quantum measurement is not a reversible operation. This means that the hidden quantum information will never be recovered (i.e, we will not be able to get back the superposed state (1)). 

We believe that the irreversibility of the standard quantum measurement is strictly related to the classical background geometry. To see why, let us consider the Bloch sphere, which is the sphere 
$S^2$ with unit radius.
Thus, any 1- qubit state can be visualized as a point on the Bloch sphere, the two basis states $\ket{0}$ and $\ket{1}$ being the poles. A standard quantum measurement of one qubit is then equivalent to the projection of one of the poles of the Bloch sphere, resulting in one point in
 $\mathbf{R}^3$, where the external observer is placed. 

Now, we wish to remind that any transformation on a qubit during a computational process is a reversible operation, as it is performed by a unitary operator $U$ such that $U^{\dag}U=\mathbf{I}$. This can be seen geometrically as follows. Any unitary matrix $U_2$ on $\mathbf{C}^2$, (which is an element of the group SU(2) multiplied by a global phase factor): 
\begin{equation}
    U_2=e^{i\phi}\left(\begin{array}{cc}
    \alpha & \beta
    \\
    -\beta^* & \alpha^*
    \end{array}\right)
\end{equation} 
(where $\alpha^*$ is the complex conjugate of $\alpha$ and $|\alpha|^2+|\beta|^2=1$), can be rewritten in terms of a rotation of the Bloch sphere: $U_2=e^{i\phi}R_n(\theta)$, where $R_n(\theta)$ is the rotation matrix of the Bloch sphere by an angle $\theta$ about an axis $n$. 

However, a projector is not a unitary operator, and it cannot be rewritten in terms of a rotation of the Bloch sphere. This means that the observer who has performed the standard quantum measurement, is not able to recover the original state by a rotation of the Bloch sphere. In fact, what the external observer sees, is just one pole of the Bloch sphere. 
The question is now whether a reversible quantum measurement could be feasible, at least in principle. Of course, the projector should be replaced by a unitary operator, but this means that the reversible measurement should be performed ''inside'' the quantum computer. Or, in other words, the hypotetical observer should be placed in a quantum space whose states are in a one-to-one correspondence with the quantum computational states, that is, a discrete quantum space associated with the algebra of quantum logic gates. Now, n-dimensional quantum logic gates are $n\times n$ unitary matrices with $n=2^N$, where $N$ is the number of qubits in the quantum register: for example, in the case of one qubit, the quantum logic gates are $2\times 2$ unitary matrices. Thus, quantum logic gates form a subset of the set of $n\times n$ complex matrices, whose algebra is a non-commutative $C^*$-algebra \cite{ped}. To this algebra, it is associated (by the non-commutative version of the Gelfand-Naimark theorem \cite{wor}), a quantum space which is the fuzzy sphere with $n$ elementary cells. This means that the computational state of a quantum computer with $N$ qubits, can be geometrically viewed as a fuzzy sphere with $n=2^N$ cells. 

We recall here that the fuzzy sphere is constructed replacing the algebra of polynomials on the (unit) sphere $S^2$ by the non commutative algebra of complex $n\times n$ matrices, which is obtained by quantizing the coordinates $x_i$ $(i=1,2,3)$:
\begin{equation}
    x_i\to X_i=kJ_i
\end{equation} 
where the $J_i$ form the $n$-dimensional irreducible representation of the algebra of SU(2),
and the non-commutativity parameter $k$ is, for a unit radius $k=1/\sqrt{n^2-1}$.
Then, the ensemble of all rotations of the Bloch sphere can be viewed geometrically as a fuzzy sphere in the $n=2$ case, where the $x_i$ are replaced by: $x_i\to X_i=\sigma_i/\sqrt{3}$, and the $\sigma_i$ are the Pauli matrices.

In what follows, we will generalize the standard quantum measurement of one qubit by using $2\times 2$ unitary matrices and we will analize the associated geometries.

To start, let us consider the diagonal $2\times 2$ matrices on the complex numbers (they form a commutative $C^*$-algebra, which is a subalgebra of the non-commutative $C^*$-algebra of complex $2\times 2$ matrices). Recall, however, that we shall require unitarity, so that we should consider only diagonal $2\times 2$ matrices of the kind: 
\begin{equation}
    U_2^D=e^{i\phi}\left(\begin{array}{cc}
    \alpha & 0
    \\
    0 & \alpha^*
    \end{array}\right)
\end{equation} 
with $|\alpha|^2$ where $U_2^D$ in (4) is the particular case of $U_2$ in (2) with $\beta=0$. The associated space to this algebra, is a 2- points lattice, that is a subspace \cite{mar} of the fuzzy sphere.

The action of $U_2^D$ on the qubit state (1) gives:
\begin{equation}
    U_2^D\ket{\psi}=a'\ket{0}+b'\ket{1}
\end{equation} 
with $a'=e^{i\phi}\alpha a$, $b'=e^{i\phi}\alpha^* b$, and: $|a'|^2=|a|^2$, $|b'|^2=|b|^2$. 
That is, the probabilities are unchanged.
Notice that, geometrically, this is equivalent to project both the poles of the Bloch sphere at the same time. The associated space is a 2-points lattice.
Now, $U_2^D$ in (4) can be rewritten as:
\begin{equation}
    U_2^D\ket{\psi}=e^{i\phi}\left(\alpha P_0+\alpha^* P_1\right)
\end{equation} 
which is a linear superposition of the two projectors $P_0$ and $P_1$: this the reversible origin of a standard (irreversible) quantum measurement in the computational basis. The application of $U_2^D$ to the state $\ket{\psi}$ in (1) is a superposition of two standard quantum measurements made at the same time. We will call this new kind of quantum measurement {\it basic} measurement in the computational basis.

After the basic measurement, the state $\ket{\psi}$ is left in the state:
\begin{equation}
    \ket{\psi}\to\ket{\psi'}=\frac{
    U_2^D\ket{\psi}}{\sqrt{|a'|^2+|b'|^2}}=e^{i\phi}\left(\alpha a\ket{0}+\alpha^* b\ket{1}\right)
\end{equation} 
where, in (7), the total probability has been considered.

The state $\ket{\psi'}$ is still a superposed state, and from it it is possible to recover the original state $\ket{\psi}$ by performing the inverse operation:
\begin{equation}
    \left(U_2^D\right)^{-1}\ket{\psi'}=\ket{\psi}
\end{equation} 
It should be noticed that the internal observer uses the projectors $P_0$ and $P_1$ at the same time. If an external observer should try to do the same, she would fail, as she lives in a classical, continuous space, namely, $\mathbf\mathbf\mathbf{R}^3$. Or, she could try to achieve the same result obtained by the internal observer by using first $P_0$ and then $P_1$ on the same state $\ket{\psi}$, but that is forbidden by the no-cloning theorem \cite{woo}, by which an unknown quantum state cannot be copied. Then, the only thing that the external observer can do, is to use either $P_0$ or $P_1$, that is, to perform a standard quantum measurement. 

In passing from the basic measurement to the standard quantum measurement, the associated geometry has changed: from the 2-points lattice to a single point.

In the context of the basic measurement, it should be noticed, however, that the 2-points lattice breaks SO(3) invariance, so that, from this space, the internal observer cannot reach any other 1-qubit state of the Bloch sphere, by a generic rotation. She can just make a quite limited operation: a rotation about the z-axis.
The logical interpretation of the basic measurement scheme is under study \cite{bat}.

Of course, it is also possible to perform a basic measurement in a different basis, for example in the dual basis.

When instead of considering the diagonal $2\times 2$ unitary matrices in (4) one considers the $2\times 2$ unitary matrices in (2), the two points of the lattice are replaced by two cells of a fuzzy sphere. Algebraically, the original qubit in (1) has been rotated into another qubit, so that its original probability amplitudes have been ''mixed up''.
In fact, the action of (3) on the qubit state (1) gives:
\begin{equation}
    U_2\ket{\psi}=e^{i\phi}
    \left(\begin{array}{cc} 
    \alpha & \beta\\
    -\beta^* & \alpha^*
    \end{array}\right)
     \left(\begin{array}{c} 
    a\\b
    \end{array}\right)
    =\left(\begin{array}{c}
    \alpha a+\beta b\\-\beta^* a+ \alpha^* b\end{array}\right)=\ket{\psi'}
\end{equation} 
The transformation in (12) does not conserve the original probabilities. 

To summarize, the hypothetical internal observer should place herself in a fuzzy sphere if she wishes to follow the whole computational process from inside the quantum computer, but she can just stand on a subspace of the fuzzy sphere, namely, the 2-points lattice, if she wants to perform a basic measurement.

Finally, we wish to say that, in this paper, the hypothetical insider observer had just the role to indicate the way for a logical insight into the black box-like quantum computational state, and to show how the standard quantum measurement originates from a reversible operation. 

\section*{Aknowledgments}
I am very grateful to Giulia Battilotti for many useful discussions.

Work partially supported by the research project ''Logical Tools for Quantum Information Theory'', Department of Mathematics, University of Padova, Italy.

 
\end{document}